\documentclass[a4]{revtex4}
\usepackage[utf8]{inputenc} %
\usepackage{amssymb}
\usepackage{graphicx}

\begin{document}

\title{Kissing loop interaction in adenine riboswitch: insights from umbrella sampling simulations}

\author{%
Francesco di Palma,
Sandro Bottaro, and
Giovanni Bussi
\footnote{To whom correspondence should be addressed.
Email: bussi@sissa.it
}}

\affiliation{%
Scuola Internazionale Superiore di Studi Avanzati, International School for Advanced Studies,
265, Via Bonomea I-34136 Trieste, Italy
}

\begin{abstract} %
{\bf Introduction} %
Riboswitches are cis-acting regulatory RNA elements prevalently located in the leader sequences of bacterial mRNA.
An adenine sensing riboswitch cis-regulates adeninosine deaminase gene ({\it add}) in {\it Vibrio vulnificus}.
The structural mechanism regulating its conformational changes upon ligand binding mostly remains to be elucidated.
In this open framework it has been suggested that the ligand stabilizes the interaction of the distal ``kissing loop'' complex.
Using accurate full-atom molecular dynamics with explicit solvent in combination with enhanced sampling techniques and advanced analysis methods
it could be possible to provide a more detailed perspective on the formation of these tertiary contacts.

{\bf Methods} %
In this work, we used umbrella sampling simulations to study the thermodynamics of the kissing loop complex in the presence and in the absence of the cognate ligand.
We enforced the breaking/formation of the loop-loop interaction restraining the distance between the two loops.
We also assessed the convergence of the results by using two alternative initialization protocols.
A structural analysis was performed using a novel approach to analyze base contacts.

{\bf Results} %
Contacts between the two loops were progressively lost when larger inter-loop distances were enforced.
Inter-loop Watson-Crick contacts survived at larger separation when compared with non-canonical
pairing and stacking interactions. Intra-loop stacking contacts remained formed upon loop undocking.
Our simulations qualitatively indicated that the ligand could stabilize the kissing loop complex.
We also compared with previously published simulation studies.

{\bf Discussion and Conclusions} %
Kissing complex stabilization given by the ligand was compatible with available experimental data.
However, the dependence of its value on the initialization protocol of the umbrella sampling simulations
posed some questions on the quantitative interpretation of the results and 
called for better converged enhanced sampling simulations.

\end{abstract}

\maketitle

\section*{Introduction}
Riboswitches are portions of ribonucleic acid (RNA) able to regulate gene expression in bacteria and plants at several levels. They bind their sensed ligands without the need for protein factors. To regulate their target gene, riboswitch can either act on transcription, on translation, or, more rarely, as interfering, antisense or self-splicing RNAs \cite{breaker2012riboswitches}. More precisely, riboswitches are cis-acting RNA elements prevalently located in the leader sequences of bacterial mRNA \cite{serganov2013decade} that regulate the expression of the same gene from which they have been transcribed. They are composed of an aptamer domain that binds the effector ligand and of an expression platform, usually located downstream of the aptamer, that transduces the ligand-induced conformational switch into the gene expression regulation \cite{barrick2007distributions,garst2009switch}. Riboswitches are classified according to the nature of the sensed ligand \cite{breaker2012riboswitches}. Among them, the purine-sensing riboswitches emerge as important model systems for exploring various aspects of RNA structure and function \cite{porter2014purine} because of their structural simplicity and relatively small size. Within the purine family the {\it add} adenine-sensing riboswitch (A-riboswitch) is one of the most characterized. Found in the mRNA 5'-untraslated region, it cis-regulates the adenosine deaminase gene in {\it Vibrio vulnificus} acting rho-independently at the translational level \cite{mandal2004adenine}. Its regulatory activity depends on the availability of the ligand: in the presence of adenine the riboswitch is in the ON state, and the protein synthesis is permitted, whereas in the absence of the ligand the riboswitch folds into the OFF state blocking the translation initiation (Figure 1). The ligand-bound structure of its aptamer \cite{serganov2004structural,mandal2004gene} is a junction of three stems (P1, P2, P3) with the ligand completely encapsulated into the structure (Figure 2). There are three structurally important regions: the binding pocket, the P1-stem and the loop-loop tertiary interaction between L2 and L3, usually called ``kissing loops''. The latter includes two inter-loop Watson-Crick (WC) base pairs \cite{forsdyke1995stem,nowakowski1997rna}. The ligand-dependent structural mechanism inducing the switch between the ON- and the OFF-state in the A-riboswitch mostly remains to be elucidated. The role of the ligand in the structural organization of the aptamer has been investigated using structure-based fluorescence spectroscopy \cite{rieder2007ligand}, multidimensional NMR techniques \cite{lee2010real} and single-molecule experiments \cite{neupane2011single}. These investigations however lack both the atomistic details and the distinct energetic contributions associated to ligand binding. In this open framework, in particular, it has been suggested experimentally that the ligand stabilizes the interaction of the distal kissing complex \cite{leipply2011effects}. At the same time a stable kissing interaction seems to contribute to the ligand binding energy stabilizing the complex in a cooperative fashion \cite{porter2014purine,rieder2007ligand,lee2010real}. Also {\it in silico} techniques have been used obtaining an accurate description of the system from a structural point of view \cite{sharma2009md,priyakumar2010role,gong2011role,lin2014sequence}.
  \begin{figure}
  \includegraphics[width=0.5\textwidth]{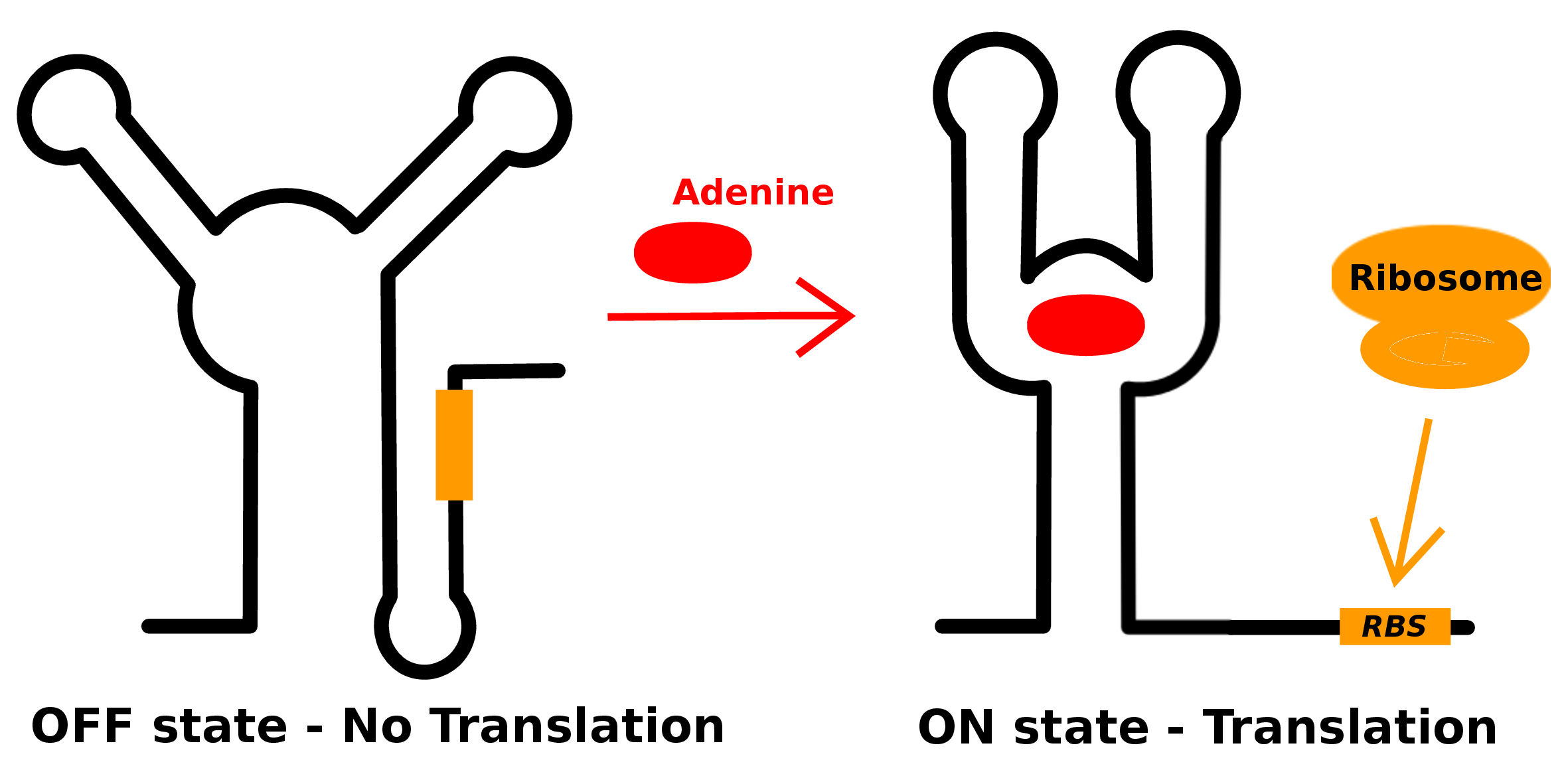}
  \caption{{\bf Mechanism of action of A-riboswitch}
     Cartoon showing the OFF (left) and ON (right) states of the A-riboswitch.
     When ligand is not present the ribosome binding site (orange, RBS) is
     paired with a portion of the aptamer and translation is blocked. When ligand
     is present the RBS is free to interact with the ribosome and translation
     can be initiated.
     }
   \end{figure}
\begin{figure}
  \includegraphics[width=0.5\textwidth]{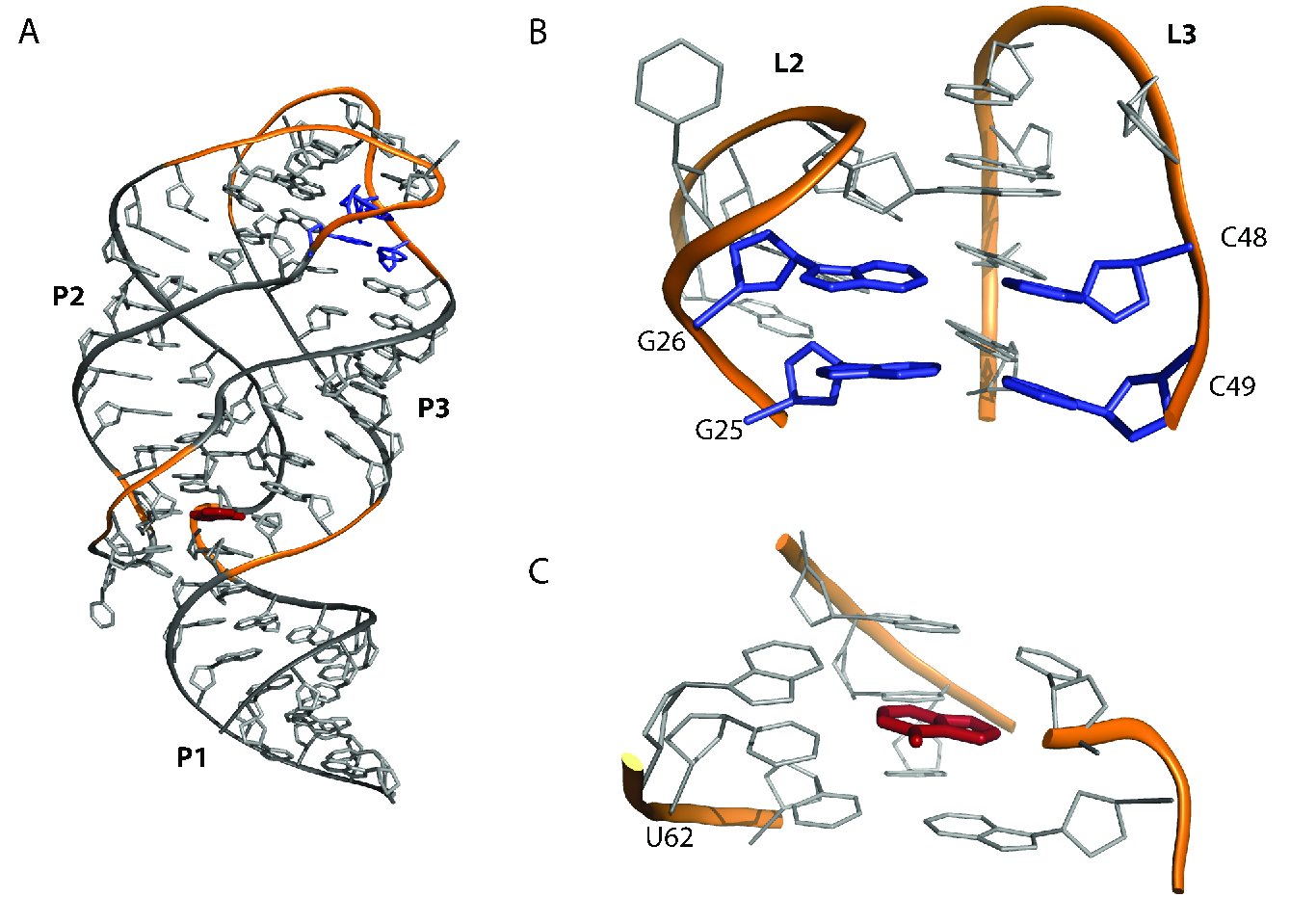}
  \caption{{\bf Structure of the {\it add} aptamer domain}
      A) Three dimensional representation of the aptamer with the adenine bound.
       The stems are shown in grey and labeled. The backbone of the loops and of the junctions is shown in orange.
      B) Close-up on the loop-loop (L2 and L3) interaction with focus on the Watson-Crick base pairs (G25-C49, G26-C48, in blue).
      C) Close-up on the ligand binding site with the adenine (red) paired with U62.
      }
      \end{figure}

In a few cases a computational approach has been employed to provide a thermodynamic characterization of the system \cite{lin2008relative,allner2013loop,di2013ligand}.
In particular, Allner {\it et al.} \cite{allner2013loop}
computed the free-energy profile corresponding to the formation of the kissing complex
using molecular dynamics (MD) simulations in explicit solvent,
both in the presence and in the absence of the ligand, using the CHARMM36 force field \cite{foloppe2000all,mackerell2000all}.
MD does not require experimental inputs and can in principle be used in a predictive fashion. However, accuracy of atomistic force fields is still debated and it is thus very important to compare results obtained employing different sets of parameters.

In this paper we use atomistic MD with the latest variant of the Amber force field
\cite{zgarbova2011refinement} in combination with enhanced sampling techniques
 \cite{abrams2013enhanced} to provide a more detailed perspective on the formation of the kissing loop complex.
The combined approach allows this contribution to be dissected from the other ligand-aptamer interactions and the impact of the ligand on the stability of the loop-loop interaction to be quantified.
We reproduce exactly the same protocol that has been used by Allner {\it et al.} \cite{allner2013loop}
in order to perform a fair comparison between the two force-fields on this particular system.
Effects of the initialization protocol on the results of umbrella sampling simulations are also discussed in detail.

\section*{Methods}
In this work, we used umbrella sampling (US) simulations \cite{torrie1977nonphysical} to study the thermodynamics of the kissing loop in the presence and in the absence of the cognate ligand. We enforced the breaking/formation of the loop-loop interaction steering the distance between the two loops and then used the resulting structures as starting conformations for US with multiple restraints \cite{kumar1992weighted}. Simulations were carried out with the Gromacs 4.6.3 program package \cite{pronk2013gromacs} combined with the PLUMED 2.0 plug-in \cite{tribello2014plumed}. All the simulation parameters are discussed in detail in the following subsections.

\subsection*{System set up}
All simulations reported hereafter were performed on two systems: the {\it add} aptamer domain complexed with adenine (Holo form) and without adenine bound (Apo form). In both cases we used the X-ray structure solved by Serganov {\it et al.} [PDB:1Y26] \cite{serganov2004structural}. The ligand was removed to simulate the unbound state. MD simulations were performed using the Amber99 force field \cite{wang2000well} refined with the parmbsc0 $\alpha$/$\gamma$ corrections \cite{perez2007refinement} and the latest $\chi$ torsional parameters \cite{zgarbova2011refinement}. The general Amber force field \cite{wang2004development} was used to parametrize the ligand. Partial atomic charges were assigned using the restricted electrostatic potential fit method \cite{bayly1993well} based on an electronic structure calculation at the HF/6-31G* level of theory performed with Gaussian03 \cite{gaussian03}. The electrostatic interactions were calculated using the particle-mesh Ewald method \cite{darden1993particle} and bond-lengths were constrained with LINCS \cite{hess1997lincs}. The systems were set-up following exactly the protocol described in Alln\'er {\it et al.} \cite{allner2013loop}: aptamers were solvated in a rhombic dodecahedron having 8 nm as box vector lenghts, with a Mg$^{2+}$-H$_2$O solution using approximately 11000 TIP3 molecules \cite{jorgensen1983comparison}, and a recent parametrization for divalent cations \cite{allner2012magnesium}. The 5 crystallographic Mg$^{2+}$ were initially kept at their respective position, whereas the additional 30 ions added to neutralize the system ([Mg$^{2+}$] = 0.18 M) were randomly placed. A steepest descent minimization (150 steps) was performed followed by 200 ps of MD at constant temperature (298 K, using stochastic velocity rescaling \cite{bussi2007canonical}) and pressure (1 atm, using the Berendsen barostat \cite{berendsen1984molecular}) with positional restraints on both RNA and ions so as to equilibrate water. This procedure was repeated first removing the constraints on the ions and then removing all the remaining constraints. Finally, 12 ns unrestrained simulations at constant volume were performed for each system.

\subsection*{Umbrella sampling}
In order to compute the thermodynamic stability of the loop-loop interaction we employed US simulations with multiple harmonic restraints. The distance between the center of mass (CoM) of the backbone atoms of the two loops (L2: bases 20-26; L3: bases 48-54; Figure 2B) was used as a collective variable (CV). We will refer hereafter to this distance as {\it L}. 44 uniformly spaced reference values were taken in the range spanning from 12.5 to 34 \AA, and restraints with stiffness
$k = 20~(kcal/mol)/$\AA$^2$ were employed. In the production phase of the US simulations each of the 44 windows was run for 5 ns.
A very important issue in US simulation is the generation of the starting conformations. We here performed two independent US sampling simulations, using starting conformations generated with two different protocols (hereafter referred to as {\it forward} and {\it backward}).
To generate the starting points for the forward US simulations we employed the same protocol used by Alln\'er {\it et al.} \cite{allner2013loop}. Namely, we ran a series of 44 short (0.25 ns) simulations with a stiffer restraint ($k = 40~(kcal/mol)/$\AA$^2$) keeping the CV at the 44 reference values, where each simulation was initialized from the last frame of the previous one. In this way, before each US window starting structure was sampled, we let the system equilibrate. The reference values were iterated allowing an increasing distance between the loops.
The backward US simulation was initialized with an equivalent procedure but iterating the restraints in the opposite order, i.e. starting from the structure with undocked loops. In principle, if US simulations are converged, the result should be independent of the initialization procedure.
 
\subsection*{Analysis methods}
The data were analyzed using the last 4 ns of each window. The potential of mean force (PMF) profiles were constructed using the weighted histogram analysis method (WHAM) \cite{kumar1992weighted} implemented by Grossfield \cite{gross209} taking the CV values distribution resulting from the US simulations.
This implementation of WHAM allows to compute errors with a bootstrapping procedure that assumes
uncorrelated samples. To avoid artifacts due to possible correlations we instead adopted a blocking procedure.
Namely, we split the final 4 ns of each trajectory in four blocks of 1 ns each and performed the WHAM calculation
using only a single block from each simulations. The four resulting profiles are aligned at their CV starting value
(12.1 \AA{} for the forward profiles, 34.4 \AA{} for the backward profiles) and error at each point is computed as the standard deviation among the four profiles divided by $\sqrt{4}$.

To define the number of stacking interactions and the number of base-pair contacts a local coordinate system was constructed in the centre of each six-membered rings, with the $x$ axis pointing towards the C2 atom and the $z$ axis orthogonal to the ring plane. The pairing and stacking relationship between two bases {\it j} and {\it k} is based on the vector {\it $r_{jk}$}, i.e. the position of ring center {\it k} relative to the coordinate system constructed on base {\it j}.
The criteria for determining the canonical WC base pairs are: 1) the base pair must be AU or GC; 2) The relative position of the bases is compatible with the geometry of a WC interaction. The latter condition is considered satisfied when the product of the Gaussian function $\mathcal{N}(r_{jk};\mu,\sigma)\times\mathcal{N}(r_{kj};\mu,\sigma) > 10^{-8}$. Mean $\mu$ and covariance $\sigma$ were obtained from the empirical distribution of WC pairs in the crystal structure of the large ribosomal subunit \cite{klein2004roles}. The criteria for determining the non-canonical base pairs are: 1) the ellipsoidal distance $\mathcal{D}_{jk} \equiv \sqrt{x_{jk}^2/25 + y_{jk}^2/25 + z_{jk}^2/9}  < \sqrt{2.5}$, and $\mathcal{D}_{kj} < \sqrt{2.5}$; 2) $|z_{jk}|$ and $|z_{kj}| < 2$~\AA; 3) it is not a WC pair. The criteria for determining the stacking base pairs are: 1) the ellipsoidal distance $\mathcal{D}_{jk}  < \sqrt{2.5}$, and $\mathcal{D}_{kj} <  \sqrt{2.5}$; 2) $|z_{jk}|$ and $|z_{kj}| > 2$~\AA; 3) $x_{jk}^2 + y_{jk}^2 < 5$~\AA{} and $x_{kj}^2 + y_{kj}^2 < 5$~\AA. This procedure yields similar results compared to the MC-annotate software \cite{gendron2001quantitative} and was shown to be useful for characterizing both structural and dynamical properties of RNA molecules \cite{bottaro14nar}. 
The software used to perform this structural analysis is available online (http://github.com/srnas/barnaba).

\section*{Results}

\subsection*{Forward process}
\begin{figure}
  \includegraphics[width=0.5\textwidth]{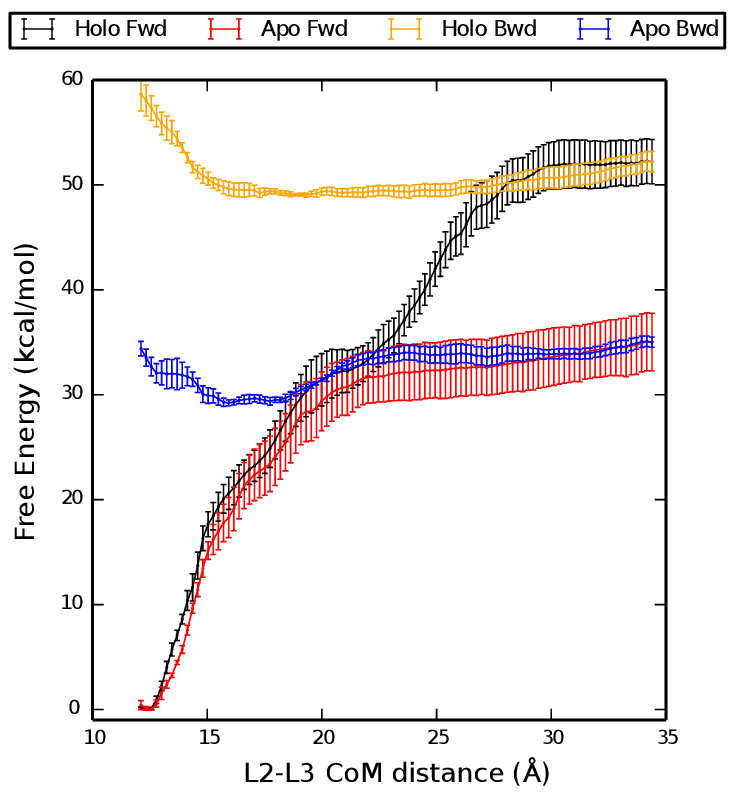}
  \caption{{\bf Potential of mean force for kissing-complex formation}
Potential of mean force (PMF) as a function of the distance between the centers
of mass of the L2 and L3 loops. Results for Holo and Apo forms are shown as
obtained from two independent umbrella sampling simulations using different
protocols to obtain the initial structures (forward, Fwd, and backward, Bwd, see main text for definition).
Fwd and Bwd profiles are aligned at the maximum distance.
}
\end{figure}

The analysis of the Holo forward and Apo forward US trajectories allowed the PMF for the disruption of the kissing complex to be computed. The resulting profiles are plotted in Figure 3 for both Holo and Apo systems.
The PMF shows a minimum at $L\approx$ 12.5 \AA, corresponding to the initial structure. The free energy change upon disruption of the kissing complex for the Holo structure is $\Delta G$ = 52 $\pm$ 2 kcal/mol. For the Apo structure the stability of the complex is largely reduced to $\Delta G$ = 35 $\pm$ 3 kcal/mol. The stabilization of the kissing complex provided by the ligand can thus be estimated as $\Delta\Delta G$ = 17 $\pm$ 3 kcal/mol.
To understand which are the interactions that are relevant for the kissing complex formation we analyze inter-loop pairings and inter- and intra-loop stacking interactions for each of the restrained simulations (Figure 4).
      \begin{figure}
  \includegraphics[width=0.5\textwidth]{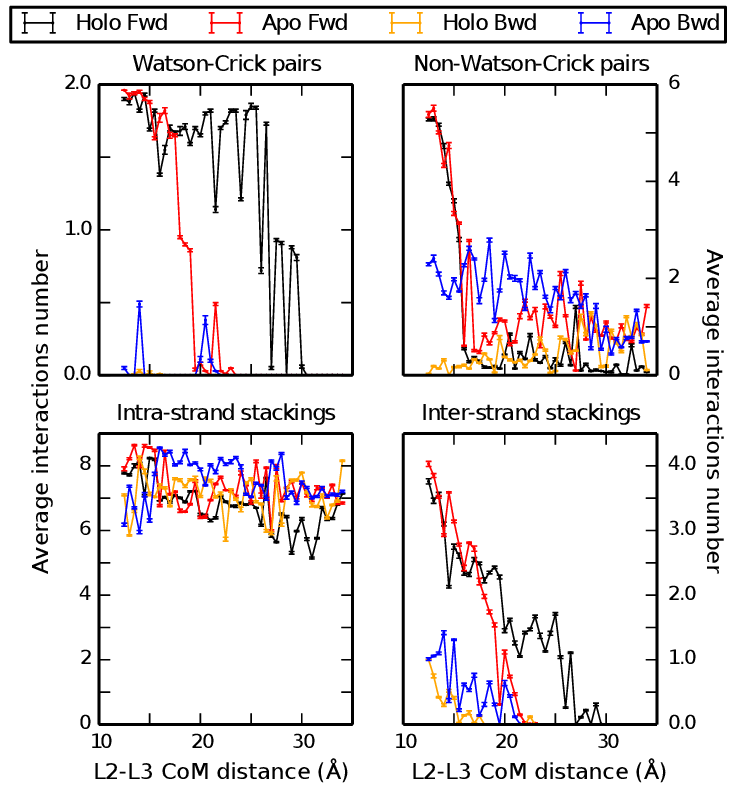}
  \caption{{\bf Analysis of inter and intra-loop interactions}
  Average count of inter and intra-loop interactions from umbrella sampling
  simulations. Results are shown for both Apo and Holo forms, using
  both forward (Fwd) and backward (Bwd) protocols (see main text for definition).
  Watson-Crick and non-Watson-Crick pairings as well as intra and inter-loop stackings
  are shown as indicated.
}
      \end{figure}

For both the Holo and the Apo forms, at a $L\gtrapprox 16$ \AA{}, only the two inter-loop WC base pairs (G25-C49, G26-C48) peculiar of the loop-loop interaction are still formed.
On the contrary, all the non-canonical base pairs are disrupted.
In the Apo structure the inter-loop WC pairings were irreversibly lost at $L > 23$ \AA, whereas in the Holo structure they are at least partially maintained until $L \approx 30$ \AA.
We also analyzed the rupture of stacking interactions, distinguishing intra-loop and inter-loop contacts.
Inter-loop stacking behaves in a manner qualitatively similar to the inter-loop WC pairings, going to zero at a distance $L \approx 23$ \AA{} (Apo) and 30 \AA{} (Holo). On the contrary, the intra-loop stacking interactions are still present when the kissing loop is disrupted, indicating that the internal structure of the two loops is preserved during undocking.
It can be observed that in the Holo simulation the number of intra-strand stacking slightly decreases ($\approx 5$) for 29 \AA{} $\lesssim L \lesssim 30$ \AA{} because of the distortion in the structure induced by the one of the two inter-loop WC pairings. After this residual interaction is lost all the intra-loop stacking contacts are recovered.

\subsection*{Backward process}
In order to better assess the convergence of the free energy landscape for kissing complex formation, we also reconstructed the PMF profiles of the Holo and Apo structure from US simulations initialized with the backward process (Figure 2). The forward and backward profiles were aligned at $L = 34$ \AA, since at that distance the starting structure of the backward process is equal to the final structure of the forward one. The free-energy change upon docking is estimated taking the difference between the minimum value of the PMF (Holo: $L\approx 19$ \AA; Apo: $L\approx 16$ \AA) and its value for the undocked structure (L = 34 \AA): for the Holo $\Delta G$ = -3.2 $\pm$ 0.9 kcal/mol, for the Apo $\Delta G$ = -5.9 $\pm$ 0.6 kcal/mol. Albeit negative, these numbers are too small and not compatible with the ones found in the forward process. This is a clear signature of hysteresis in the pulling procedure that strongly biases the initial starting points of the US simulation. The reason for this discrepancy can be better understood by performing a structural analysis of the interactions on the different US windows. As it can be seen in Figure 4, in the backward process the native WC base pairs are not reformed. In general, a few contacts are formed between the two loops but they are not enough to stabilize the kissing complex.
To be sure that this is a systematic effect we also tried a few alternative settings for backward simulations. Results are presented in the Appendix.

\section*{Discussion and Conclusions}

Our calculations provide quantitative and atomistic details on the mechanism of kissing loop breaking and formation in the {\it add} riboswitch aptamer domain.
The results can be compared with those recently obtained by Alln\'er {\it et al.} \cite{allner2013loop} on the same system using the CHARMM36 force field \cite{foloppe2000all,mackerell2000all}.
In particular the free-energy computed with the forward process has been obtained with an identical protocol
so as to allow a fair comparison between the force fields.
In our work the estimated stability of the kissing loop complex is $\Delta G$ = 52 $\pm$ 2 kcal/mol (Holo) and $\Delta G$ = 35 $\pm$ 3 kcal/mol (Apo),
so that upon ligand binding $\Delta \Delta G$=17 kcal/mol $>0$.
On the contrary, Alln\'er {\it et al.} reported $\Delta \Delta G=-10$ kcal/mol $<0$.
The sign of $\Delta \Delta G$ indicates whether the ligand binding and the formation of the kissing loop complex are cooperative (positive) or anticooperative (negative). Results obtained with the two force fields thus interestingly lead to two opposite pictures.

Recent experiments probed the differential ligand affinity in aptamers with mutations hindering the formation of the kissing complex \cite{leipply2011effects}. The change in affinity indicates a cooperativity
between ligand binding and kissing complex formation. This stabilization has been estimated to be $\Delta \Delta G \approx 6$ kcal/mol. This number should be interpreted with caution since it is based on the assumption that the mutated aptamer mimics a ligand-bound state that is accessible to the wild type aptamer \cite{leipply2011effects}. Results obtained with Amber force field are in qualitative agreement with this picture.

Recently, the thermodynamics of other stand-alone kissing complexes has been also studied using different biophysical techniques \cite{salim2012thermodynamic,stephenson2013essential}. In these two experimental works the stability of the loop-loop interactions resulted to be in the range 8 $\lesssim \Delta G \lesssim$ 14 kcal/mol.
Stability depends on the exact sequence and set of intra-loop interactions, but is always on the order of ten kcal/mol.
The estimated stability of the kissing loop complex in our calculation,
namely  $\Delta G$ = 52 $\pm$ 2 kcal/mol and $\Delta G$ = 35 $\pm$ 3 kcal/mol for Holo and Apo respectively, is thus much larger than expected.
Results obtained with CHARMM indicate a lower $\Delta G$ for both the systems \cite{allner2013loop}, in better agreement with experimental results, even if still overestimated.
Our result could be affected by the known overestimation of stacking interactions in the Amber force field \cite{banas2012can}.
Additionally, we would like to point out that the overestimation found in our calculations could also be a consequence of difficult convergence in the US simulations.
To test if the US simulation are effectively converged, we tried to recover the profiles from simulations initialized with the backward process, with a procedure inspired by two directional pulling in steered MD \cite{grubmuller1996ligand,minh2008optimized,do2013rna}.
The discrepancy between forward and backward process is an index of high dependence of the PMF on the initialization procedure and poses some questions on the actual convergence of
the US simulations. 
Similarly to steered MD, one can expect that simulations in the forward and backward process are respectively overestimating and underestimating the kissing complex stability.
Optimal results can be obtained in steered MD by combining simulations performed with both protocols \cite{minh2008optimized,do2013rna}.
We stress here that even if the forward simulations apparently recover the qualitative behavior of the general accepted model, they cannot be trusted for a quantitative estimation of the free-energy change. The fact that the backward process cannot reach the native docked state is a signature of a barrier in an orthogonal degree of freedom that is not properly sampled. A possible candidate is the barrier related to the desolvation of the loops, required to form the correct interstrand interactions. Additionally, we observe that pulling on the distance between the two loops does not necessarily induce the entropic reduction required upon docking. These issues are expected to affect both forward and backward pulling. Our simulation could not give an estimate of the additional barriers, but we can assume that these issues equally affects the Holo and the Apo systems. Thus, the converged $\Delta \Delta G$ upon ligand binding should be somewhere in between results from the forward and backward simulations.
Thus we can expect the $\Delta \Delta G$ to be in a wide range between -2.7 kcal/mol and 17 kcal/mol, which is in qualitative agreement with already mentioned experiments \cite{leipply2011effects}.

The convergence problem is not related to the US method itself but to the difficulty of describing such a complex docking event using a single distance as a CV. This variable is not sufficient to drive the system through the appropriate transition states. This is likely due to the existence of additional barriers on hidden degrees of freedom (e.g. solvation).
We believe that in order to reliably quantify the $\Delta \Delta G$ for this system  with US or other biased sampling methods  one should employ more complex CVs which are closer to the actual reaction coordinate.

In conclusion, in this work we addressed the formation of the kissing loop complex in the A-riboswitch aptamer by means of accurate molecular dynamics simulations in explicit solvent combined with enhanced sampling techniques in presence or absence of the cognate ligand.
Results are compatible with experiments and suggest that the ligand stabilizes the kissing loop formation. However, our results also spot some weakness of the umbrella sampling method and call for calculations performed with more advanced techniques, which will be the subject of future investigations.

\section*{Appendix}

In order to assess the backward procedure for the US method, we repeated it
for the Apo form using
a softer restraint ($k = 20~(kcal/mol)/$\AA$^2$). We performed two additional simulations:
\begin{itemize}
\item[A] Starting from the final snapshot of the forward procedure explained above
we perform a backward procedure with the softer restraint.
\item[B] We repeated both the forward and the backward procedures using the softer restraint
\end{itemize}
      \begin{figure}
  \includegraphics[width=0.5\textwidth]{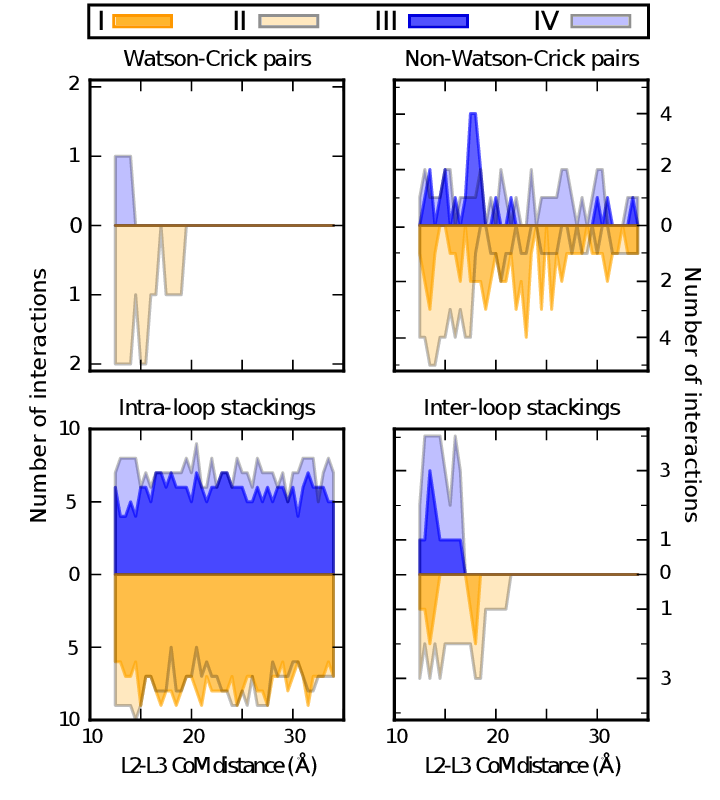}
  \caption{{\bf Analysis of interactions with alternative backward protocol}
  Count of inter and intra-loop interactions in the 44 starting snapshots resulted from the different Apo US simulation initialization procedures. Results are shown for the 4 Apo runs, using
  the backward and forward protocols, both with $k = 40 (kcal/mol)/$\AA$^2$ (I, II, respectively in orange and pale orange), and the two alternative backward protocols with the softer restraint (III in blue from protocol A, and IV in light blue from protocol B). Watson-Crick and non-Watson-Crick pairings as well as intra and inter-loop stackings are shown in the different panels.
}
      \end{figure}

Structural analysis is shown in Figure 5, where it can be appreciated that only
the simulations with protocol A (in light blue) were able to correctly form the native WC pairs.
Although the restraint stiffness could affect the result,
we believe that here the differences are mostly due to the stochastic nature of MD.

\begin{figure}
  \includegraphics[width=0.5\textwidth]{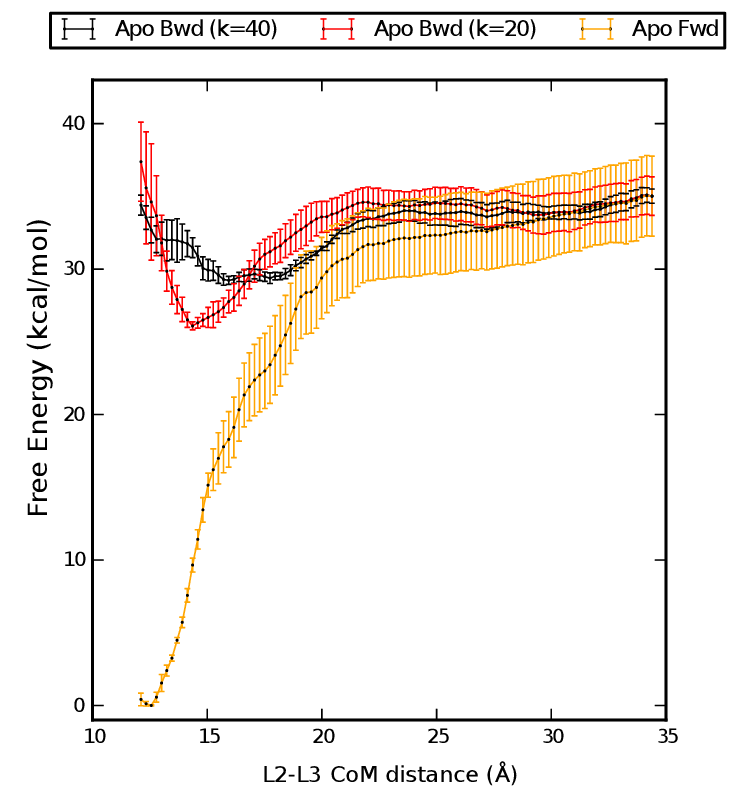}
  \caption{{\bf Potential of mean force with alternative backward protocol}
Potential of mean force (PMF) as a function of the distance between the centers
of mass of the L2 and L3 loops. Results for Apo form are shown in red (Bwd (k = 20)) as
obtained from the control umbrella sampling simulations discussed in the Appendix, and aligned with the other two Apo profiles described in the Methods section (Fwd in orange, Bwd (k = 40) in black) for comparison.
}
      \end{figure}

Using the snapshots from protocol A, we performed another US simulation. The resulting PMF profiles
are shown in Figure 6. Also in this case the free-energy landscape is incompatible
with the one obtained from the forward protocol (compare with Figure 3), indicating convergence issues in the US calculation.

\section*{Competing interests}
  The authors declare that they have no competing interests.

\section*{Author's contributions}
 FdP performed molecular dynamics simulations, analyzed the results, and drafted the manuscript.
 SB analyzed the results and helped to draft the manuscript.
 GB conceived the study and drafted the manuscript.

\section*{Acknowledgements}
Alessandra Villa is acknowledged for reading the manuscript and providing useful suggestions.

The research leading to these results has received funding from the European Research Council under the European Union's Seventh Framework
Programme (FP/2007-2013) / ERC Grant Agreement n. 306662, S-RNA-S. Publication of this article has been funded by the same ERC grant.

We acknowledge the CINECA award no. HP10BI6HS2, 2013 under the ISCRA initiative for the availability of high performance computing resources.
%
%
%
%
%
%
%
%
%
%
%
%
%
%

%

\begin{thebibliography}{50}
\expandafter\ifx\csname natexlab\endcsname\relax\def\natexlab#1{#1}\fi
\expandafter\ifx\csname bibnamefont\endcsname\relax
  \def\bibnamefont#1{#1}\fi
\expandafter\ifx\csname bibfnamefont\endcsname\relax
  \def\bibfnamefont#1{#1}\fi
\expandafter\ifx\csname citenamefont\endcsname\relax
  \def\citenamefont#1{#1}\fi
\expandafter\ifx\csname url\endcsname\relax
  \def\url#1{\texttt{#1}}\fi
\expandafter\ifx\csname urlprefix\endcsname\relax\def\urlprefix{URL }\fi
\providecommand{\bibinfo}[2]{#2}
\providecommand{\eprint}[2][]{\url{#2}}

\bibitem[{\citenamefont{Breaker}(2012)}]{breaker2012riboswitches}
\bibinfo{author}{\bibfnamefont{R.~R.} \bibnamefont{Breaker}},
  \bibinfo{journal}{Cold Spring Harbor Perspectives in Biology}
  \textbf{\bibinfo{volume}{4}}, \bibinfo{pages}{a003566}
  (\bibinfo{year}{2012}).

\bibitem[{\citenamefont{Serganov and Nudler}(2013)}]{serganov2013decade}
\bibinfo{author}{\bibfnamefont{A.}~\bibnamefont{Serganov}} \bibnamefont{and}
  \bibinfo{author}{\bibfnamefont{E.}~\bibnamefont{Nudler}},
  \bibinfo{journal}{Cell} \textbf{\bibinfo{volume}{152}}, \bibinfo{pages}{17}
  (\bibinfo{year}{2013}).

\bibitem[{\citenamefont{Barrick and Breaker}(2007)}]{barrick2007distributions}
\bibinfo{author}{\bibfnamefont{J.~E.} \bibnamefont{Barrick}} \bibnamefont{and}
  \bibinfo{author}{\bibfnamefont{R.~R.} \bibnamefont{Breaker}},
  \bibinfo{journal}{Genome Biology} \textbf{\bibinfo{volume}{8}},
  \bibinfo{pages}{R239} (\bibinfo{year}{2007}).

\bibitem[{\citenamefont{Garst and Batey}(2009)}]{garst2009switch}
\bibinfo{author}{\bibfnamefont{A.~D.} \bibnamefont{Garst}} \bibnamefont{and}
  \bibinfo{author}{\bibfnamefont{R.~T.} \bibnamefont{Batey}},
  \bibinfo{journal}{Biochimica et Biophysica Acta (BBA)-Gene Regulatory
  Mechanisms} \textbf{\bibinfo{volume}{1789}}, \bibinfo{pages}{584}
  (\bibinfo{year}{2009}).

\bibitem[{\citenamefont{Porter et~al.}(2014)\citenamefont{Porter,
  Marcano-Vel{\'a}zquez, and Batey}}]{porter2014purine}
\bibinfo{author}{\bibfnamefont{E.~B.} \bibnamefont{Porter}},
  \bibinfo{author}{\bibfnamefont{J.~G.} \bibnamefont{Marcano-Vel{\'a}zquez}},
  \bibnamefont{and} \bibinfo{author}{\bibfnamefont{R.~T.} \bibnamefont{Batey}},
  \bibinfo{journal}{Biochimica et Biophysica Acta (BBA)-Gene Regulatory
  Mechanisms}
 \textbf{\bibinfo{volume}{1839}}, \bibinfo{pages}{919}
  (\bibinfo{year}{2014}).

\bibitem[{\citenamefont{Mandal and
  Breaker}(2004{\natexlab{a}})}]{mandal2004adenine}
\bibinfo{author}{\bibfnamefont{M.}~\bibnamefont{Mandal}} \bibnamefont{and}
  \bibinfo{author}{\bibfnamefont{R.~R.} \bibnamefont{Breaker}},
  \bibinfo{journal}{Nature Structural \& Molecular Biology}
  \textbf{\bibinfo{volume}{11}}, \bibinfo{pages}{29}
  (\bibinfo{year}{2004}{\natexlab{a}}).

\bibitem[{\citenamefont{Serganov et~al.}(2004)\citenamefont{Serganov, Yuan,
  Pikovskaya, Polonskaia, Malinina, Phan, Hobartner, Micura, Breaker, and
  Patel}}]{serganov2004structural}
\bibinfo{author}{\bibfnamefont{A.}~\bibnamefont{Serganov}},
  \bibinfo{author}{\bibfnamefont{Y.-R.} \bibnamefont{Yuan}},
  \bibinfo{author}{\bibfnamefont{O.}~\bibnamefont{Pikovskaya}},
  \bibinfo{author}{\bibfnamefont{A.}~\bibnamefont{Polonskaia}},
  \bibinfo{author}{\bibfnamefont{L.}~\bibnamefont{Malinina}},
  \bibinfo{author}{\bibfnamefont{A.~T.} \bibnamefont{Phan}},
  \bibinfo{author}{\bibfnamefont{C.}~\bibnamefont{Hobartner}},
  \bibinfo{author}{\bibfnamefont{R.}~\bibnamefont{Micura}},
  \bibinfo{author}{\bibfnamefont{R.~R.} \bibnamefont{Breaker}},
  \bibnamefont{and} \bibinfo{author}{\bibfnamefont{D.~J.} \bibnamefont{Patel}},
  \bibinfo{journal}{Chemistry \& Biology} \textbf{\bibinfo{volume}{11}},
  \bibinfo{pages}{1729} (\bibinfo{year}{2004}).

\bibitem[{\citenamefont{Mandal and
  Breaker}(2004{\natexlab{b}})}]{mandal2004gene}
\bibinfo{author}{\bibfnamefont{M.}~\bibnamefont{Mandal}} \bibnamefont{and}
  \bibinfo{author}{\bibfnamefont{R.~R.} \bibnamefont{Breaker}},
  \bibinfo{journal}{Nature Reviews Molecular Cell Biology}
  \textbf{\bibinfo{volume}{5}}, \bibinfo{pages}{451}
  (\bibinfo{year}{2004}{\natexlab{b}}).

\bibitem[{\citenamefont{Forsdyke}(1995)}]{forsdyke1995stem}
\bibinfo{author}{\bibfnamefont{D.}~\bibnamefont{Forsdyke}},
  \bibinfo{journal}{Molecular Biology and Evolution}
  \textbf{\bibinfo{volume}{12}}, \bibinfo{pages}{949} (\bibinfo{year}{1995}).

\bibitem[{\citenamefont{Nowakowski and Tinoco~Jr}(1997)}]{nowakowski1997rna}
\bibinfo{author}{\bibfnamefont{J.}~\bibnamefont{Nowakowski}} \bibnamefont{and}
  \bibinfo{author}{\bibfnamefont{I.}~\bibnamefont{Tinoco~Jr}}, in
  \emph{\bibinfo{booktitle}{Seminars in Virology}} (\bibinfo{year}{1997}),
  vol.~\bibinfo{volume}{8}, pp. \bibinfo{pages}{153--165}.

\bibitem[{\citenamefont{Rieder et~al.}(2007)\citenamefont{Rieder, Lang, Graber,
  and Micura}}]{rieder2007ligand}
\bibinfo{author}{\bibfnamefont{R.}~\bibnamefont{Rieder}},
  \bibinfo{author}{\bibfnamefont{K.}~\bibnamefont{Lang}},
  \bibinfo{author}{\bibfnamefont{D.}~\bibnamefont{Graber}}, \bibnamefont{and}
  \bibinfo{author}{\bibfnamefont{R.}~\bibnamefont{Micura}},
  \bibinfo{journal}{Chembiochem} \textbf{\bibinfo{volume}{8}},
  \bibinfo{pages}{896} (\bibinfo{year}{2007}).

\bibitem[{\citenamefont{Lee et~al.}(2010)\citenamefont{Lee, Gal, Frydman, and
  Varani}}]{lee2010real}
\bibinfo{author}{\bibfnamefont{M.-K.} \bibnamefont{Lee}},
  \bibinfo{author}{\bibfnamefont{M.}~\bibnamefont{Gal}},
  \bibinfo{author}{\bibfnamefont{L.}~\bibnamefont{Frydman}}, \bibnamefont{and}
  \bibinfo{author}{\bibfnamefont{G.}~\bibnamefont{Varani}},
  \bibinfo{journal}{Proceedings of the National Academy of Sciences of the
  U.S.A.} \textbf{\bibinfo{volume}{107}}, \bibinfo{pages}{9192}
  (\bibinfo{year}{2010}).

\bibitem[{\citenamefont{Neupane et~al.}(2011)\citenamefont{Neupane, Yu, Foster,
  Wang, and Woodside}}]{neupane2011single}
\bibinfo{author}{\bibfnamefont{K.}~\bibnamefont{Neupane}},
  \bibinfo{author}{\bibfnamefont{H.}~\bibnamefont{Yu}},
  \bibinfo{author}{\bibfnamefont{D.~A.} \bibnamefont{Foster}},
  \bibinfo{author}{\bibfnamefont{F.}~\bibnamefont{Wang}}, \bibnamefont{and}
  \bibinfo{author}{\bibfnamefont{M.~T.} \bibnamefont{Woodside}},
  \bibinfo{journal}{Nucleic Acids Research} \textbf{\bibinfo{volume}{39}},
  \bibinfo{pages}{7677} (\bibinfo{year}{2011}).

\bibitem[{\citenamefont{Leipply and Draper}(2011)}]{leipply2011effects}
\bibinfo{author}{\bibfnamefont{D.}~\bibnamefont{Leipply}} \bibnamefont{and}
  \bibinfo{author}{\bibfnamefont{D.~E.} \bibnamefont{Draper}},
  \bibinfo{journal}{Biochemistry} \textbf{\bibinfo{volume}{50}},
  \bibinfo{pages}{2790} (\bibinfo{year}{2011}).

\bibitem[{\citenamefont{Sharma et~al.}(2009)\citenamefont{Sharma, Bulusu, and
  Mitra}}]{sharma2009md}
\bibinfo{author}{\bibfnamefont{M.}~\bibnamefont{Sharma}},
  \bibinfo{author}{\bibfnamefont{G.}~\bibnamefont{Bulusu}}, \bibnamefont{and}
  \bibinfo{author}{\bibfnamefont{A.}~\bibnamefont{Mitra}},
  \bibinfo{journal}{RNA} \textbf{\bibinfo{volume}{15}}, \bibinfo{pages}{1673}
  (\bibinfo{year}{2009}).

\bibitem[{\citenamefont{Priyakumar and
  Mac{K}erell~Jr}(2010)}]{priyakumar2010role}
\bibinfo{author}{\bibfnamefont{U.}~\bibnamefont{Priyakumar}} \bibnamefont{and}
  \bibinfo{author}{\bibfnamefont{A.~D.} \bibnamefont{Mac{K}erell~Jr}},
  \bibinfo{journal}{Journal of Molecular Biology}
  \textbf{\bibinfo{volume}{396}}, \bibinfo{pages}{1422} (\bibinfo{year}{2010}).

\bibitem[{\citenamefont{Gong et~al.}(2011)\citenamefont{Gong, Zhao, Chen, and
  Xiao}}]{gong2011role}
\bibinfo{author}{\bibfnamefont{Z.}~\bibnamefont{Gong}},
  \bibinfo{author}{\bibfnamefont{Y.}~\bibnamefont{Zhao}},
  \bibinfo{author}{\bibfnamefont{C.}~\bibnamefont{Chen}}, \bibnamefont{and}
  \bibinfo{author}{\bibfnamefont{Y.}~\bibnamefont{Xiao}},
  \bibinfo{journal}{Journal of Biomolecular Structure and Dynamics}
  \textbf{\bibinfo{volume}{29}}, \bibinfo{pages}{403} (\bibinfo{year}{2011}).

\bibitem[{\citenamefont{Lin et~al.}(2014)\citenamefont{Lin, Hyeon, and
  Thirumalai}}]{lin2014sequence}
\bibinfo{author}{\bibfnamefont{J.-C.} \bibnamefont{Lin}},
  \bibinfo{author}{\bibfnamefont{C.}~\bibnamefont{Hyeon}}, \bibnamefont{and}
  \bibinfo{author}{\bibfnamefont{D.}~\bibnamefont{Thirumalai}},
  \bibinfo{journal}{Phys. Chem. Chem. Phys.} \textbf{\bibinfo{volume}{16}},
  \bibinfo{pages}{6376} (\bibinfo{year}{2014}).

\bibitem[{\citenamefont{Lin and Thirumalai}(2008)}]{lin2008relative}
\bibinfo{author}{\bibfnamefont{J.-C.} \bibnamefont{Lin}} \bibnamefont{and}
  \bibinfo{author}{\bibfnamefont{D.}~\bibnamefont{Thirumalai}},
  \bibinfo{journal}{Journal of the American Chemical Society}
  \textbf{\bibinfo{volume}{130}}, \bibinfo{pages}{14080}
  (\bibinfo{year}{2008}).

\bibitem[{\citenamefont{Alln{\'e}r et~al.}(2013)\citenamefont{Alln{\'e}r,
  Nilsson, and Villa}}]{allner2013loop}
\bibinfo{author}{\bibfnamefont{O.}~\bibnamefont{Alln{\'e}r}},
  \bibinfo{author}{\bibfnamefont{L.}~\bibnamefont{Nilsson}}, \bibnamefont{and}
  \bibinfo{author}{\bibfnamefont{A.}~\bibnamefont{Villa}},
  \bibinfo{journal}{RNA} \textbf{\bibinfo{volume}{19}}, \bibinfo{pages}{916}
  (\bibinfo{year}{2013}).

\bibitem[{\citenamefont{Di~Palma et~al.}(2013)\citenamefont{Di~Palma, Colizzi,
  and Bussi}}]{di2013ligand}
\bibinfo{author}{\bibfnamefont{F.}~\bibnamefont{Di~Palma}},
  \bibinfo{author}{\bibfnamefont{F.}~\bibnamefont{Colizzi}}, \bibnamefont{and}
  \bibinfo{author}{\bibfnamefont{G.}~\bibnamefont{Bussi}},
  \bibinfo{journal}{RNA} \textbf{\bibinfo{volume}{19}}, \bibinfo{pages}{1517}
  (\bibinfo{year}{2013}).

\bibitem[{\citenamefont{Foloppe and Mac{K}erell~Jr}(2000)}]{foloppe2000all}
\bibinfo{author}{\bibfnamefont{N.}~\bibnamefont{Foloppe}} \bibnamefont{and}
  \bibinfo{author}{\bibfnamefont{A.~D.} \bibnamefont{Mac{K}erell~Jr}},
  \bibinfo{journal}{Journal of Computational Chemistry}
  \textbf{\bibinfo{volume}{21}}, \bibinfo{pages}{86} (\bibinfo{year}{2000}).

\bibitem[{\citenamefont{Mac{K}erell and Banavali}(2000)}]{mackerell2000all}
\bibinfo{author}{\bibfnamefont{A.~D.} \bibnamefont{Mac{K}erell}}
  \bibnamefont{and} \bibinfo{author}{\bibfnamefont{N.~K.}
  \bibnamefont{Banavali}}, \bibinfo{journal}{Journal of Computational
  Chemistry} \textbf{\bibinfo{volume}{21}}, \bibinfo{pages}{105}
  (\bibinfo{year}{2000}).

\bibitem[{\citenamefont{Zgarbov{\'a} et~al.}(2011)\citenamefont{Zgarbov{\'a},
  Otyepka, \v{S}poner, Ml{\'a}dek, Ban{\'a}\v{s}, Cheatham~III, and
  Jure\v{c}ka}}]{zgarbova2011refinement}
\bibinfo{author}{\bibfnamefont{M.}~\bibnamefont{Zgarbov{\'a}}},
  \bibinfo{author}{\bibfnamefont{M.}~\bibnamefont{Otyepka}},
  \bibinfo{author}{\bibfnamefont{J.}~\bibnamefont{\v{S}poner}},
  \bibinfo{author}{\bibfnamefont{A.}~\bibnamefont{Ml{\'a}dek}},
  \bibinfo{author}{\bibfnamefont{P.}~\bibnamefont{Ban{\'a}\v{s}}},
  \bibinfo{author}{\bibfnamefont{T.~E.} \bibnamefont{Cheatham~III}},
  \bibnamefont{and}
  \bibinfo{author}{\bibfnamefont{P.}~\bibnamefont{Jure\v{c}ka}},
  \bibinfo{journal}{Journal of Chemical Theory and Computation}
  \textbf{\bibinfo{volume}{7}}, \bibinfo{pages}{2886} (\bibinfo{year}{2011}).

\bibitem[{\citenamefont{Abrams and Bussi}(2013)}]{abrams2013enhanced}
\bibinfo{author}{\bibfnamefont{C.}~\bibnamefont{Abrams}} \bibnamefont{and}
  \bibinfo{author}{\bibfnamefont{G.}~\bibnamefont{Bussi}},
  \bibinfo{journal}{Entropy} \textbf{\bibinfo{volume}{16}},
  \bibinfo{pages}{163} (\bibinfo{year}{2013}).

\bibitem[{\citenamefont{Torrie and Valleau}(1977)}]{torrie1977nonphysical}
\bibinfo{author}{\bibfnamefont{G.~M.} \bibnamefont{Torrie}} \bibnamefont{and}
  \bibinfo{author}{\bibfnamefont{J.~P.} \bibnamefont{Valleau}},
  \bibinfo{journal}{Journal of Computational Physics}
  \textbf{\bibinfo{volume}{23}}, \bibinfo{pages}{187} (\bibinfo{year}{1977}).

\bibitem[{\citenamefont{Kumar et~al.}(1992)\citenamefont{Kumar, Rosenberg,
  Bouzida, Swendsen, and Kollman}}]{kumar1992weighted}
\bibinfo{author}{\bibfnamefont{S.}~\bibnamefont{Kumar}},
  \bibinfo{author}{\bibfnamefont{J.~M.} \bibnamefont{Rosenberg}},
  \bibinfo{author}{\bibfnamefont{D.}~\bibnamefont{Bouzida}},
  \bibinfo{author}{\bibfnamefont{R.~H.} \bibnamefont{Swendsen}},
  \bibnamefont{and} \bibinfo{author}{\bibfnamefont{P.~A.}
  \bibnamefont{Kollman}}, \bibinfo{journal}{Journal of Computational Chemistry}
  \textbf{\bibinfo{volume}{13}}, \bibinfo{pages}{1011} (\bibinfo{year}{1992}).

\bibitem[{\citenamefont{Pronk et~al.}(2013)\citenamefont{Pronk, P{\'a}ll,
  Schulz, Larsson, Bjelkmar, Apostolov, Shirts, Smith, Kasson, van~der Spoel
  et~al.}}]{pronk2013gromacs}
\bibinfo{author}{\bibfnamefont{S.}~\bibnamefont{Pronk}},
  \bibinfo{author}{\bibfnamefont{S.}~\bibnamefont{P{\'a}ll}},
  \bibinfo{author}{\bibfnamefont{R.}~\bibnamefont{Schulz}},
  \bibinfo{author}{\bibfnamefont{P.}~\bibnamefont{Larsson}},
  \bibinfo{author}{\bibfnamefont{P.}~\bibnamefont{Bjelkmar}},
  \bibinfo{author}{\bibfnamefont{R.}~\bibnamefont{Apostolov}},
  \bibinfo{author}{\bibfnamefont{M.~R.} \bibnamefont{Shirts}},
  \bibinfo{author}{\bibfnamefont{J.~C.} \bibnamefont{Smith}},
  \bibinfo{author}{\bibfnamefont{P.~M.} \bibnamefont{Kasson}},
  \bibinfo{author}{\bibfnamefont{D.}~\bibnamefont{van~der Spoel}},
  \bibnamefont{et~al.}, \bibinfo{journal}{Bioinformatics}
  \textbf{\bibinfo{volume}{29}}, \bibinfo{pages}{845} (\bibinfo{year}{2013}).

\bibitem[{\citenamefont{Tribello et~al.}(2014)\citenamefont{Tribello, Bonomi,
  Branduardi, Camilloni, and Bussi}}]{tribello2014plumed}
\bibinfo{author}{\bibfnamefont{G.~A.} \bibnamefont{Tribello}},
  \bibinfo{author}{\bibfnamefont{M.}~\bibnamefont{Bonomi}},
  \bibinfo{author}{\bibfnamefont{D.}~\bibnamefont{Branduardi}},
  \bibinfo{author}{\bibfnamefont{C.}~\bibnamefont{Camilloni}},
  \bibnamefont{and} \bibinfo{author}{\bibfnamefont{G.}~\bibnamefont{Bussi}},
  \bibinfo{journal}{Computer Physics Communications}
  \textbf{\bibinfo{volume}{185}}, \bibinfo{pages}{604} (\bibinfo{year}{2014}).

\bibitem[{\citenamefont{Wang et~al.}(2000)\citenamefont{Wang, Cieplak, and
  Kollman}}]{wang2000well}
\bibinfo{author}{\bibfnamefont{J.}~\bibnamefont{Wang}},
  \bibinfo{author}{\bibfnamefont{P.}~\bibnamefont{Cieplak}}, \bibnamefont{and}
  \bibinfo{author}{\bibfnamefont{P.~A.} \bibnamefont{Kollman}},
  \bibinfo{journal}{Journal of Computational Chemistry}
  \textbf{\bibinfo{volume}{21}}, \bibinfo{pages}{1049} (\bibinfo{year}{2000}).

\bibitem[{\citenamefont{P{\'e}rez et~al.}(2007)\citenamefont{P{\'e}rez,
  March{\'a}n, Svozil, Sponer, Cheatham~III, Laughton, and
  Orozco}}]{perez2007refinement}
\bibinfo{author}{\bibfnamefont{A.}~\bibnamefont{P{\'e}rez}},
  \bibinfo{author}{\bibfnamefont{I.}~\bibnamefont{March{\'a}n}},
  \bibinfo{author}{\bibfnamefont{D.}~\bibnamefont{Svozil}},
  \bibinfo{author}{\bibfnamefont{J.}~\bibnamefont{Sponer}},
  \bibinfo{author}{\bibfnamefont{T.~E.} \bibnamefont{Cheatham~III}},
  \bibinfo{author}{\bibfnamefont{C.~A.} \bibnamefont{Laughton}},
  \bibnamefont{and} \bibinfo{author}{\bibfnamefont{M.}~\bibnamefont{Orozco}},
  \bibinfo{journal}{Biophysical Journal} \textbf{\bibinfo{volume}{92}},
  \bibinfo{pages}{3817} (\bibinfo{year}{2007}).

\bibitem[{\citenamefont{Wang et~al.}(2004)\citenamefont{Wang, Wolf, Caldwell,
  Kollman, and Case}}]{wang2004development}
\bibinfo{author}{\bibfnamefont{J.}~\bibnamefont{Wang}},
  \bibinfo{author}{\bibfnamefont{R.~M.} \bibnamefont{Wolf}},
  \bibinfo{author}{\bibfnamefont{J.~W.} \bibnamefont{Caldwell}},
  \bibinfo{author}{\bibfnamefont{P.~A.} \bibnamefont{Kollman}},
  \bibnamefont{and} \bibinfo{author}{\bibfnamefont{D.~A.} \bibnamefont{Case}},
  \bibinfo{journal}{Journal of Computational Chemistry}
  \textbf{\bibinfo{volume}{25}}, \bibinfo{pages}{1157} (\bibinfo{year}{2004}).

\bibitem[{\citenamefont{Bayly et~al.}(1993)\citenamefont{Bayly, Cieplak,
  Cornell, and Kollman}}]{bayly1993well}
\bibinfo{author}{\bibfnamefont{C.~I.} \bibnamefont{Bayly}},
  \bibinfo{author}{\bibfnamefont{P.}~\bibnamefont{Cieplak}},
  \bibinfo{author}{\bibfnamefont{W.}~\bibnamefont{Cornell}}, \bibnamefont{and}
  \bibinfo{author}{\bibfnamefont{P.~A.} \bibnamefont{Kollman}},
  \bibinfo{journal}{Journal of Physical Chemistry}
  \textbf{\bibinfo{volume}{97}}, \bibinfo{pages}{10269} (\bibinfo{year}{1993}).

\bibitem[{\citenamefont{Frisch et~al.}(2004)\citenamefont{Frisch, Trucks,
  Schlegel, Scuseria, Robb, Cheeseman, Montgomery, Vreven, Kudin, Burant
  et~al.}}]{gaussian03}
\bibinfo{author}{\bibfnamefont{M.}~\bibnamefont{Frisch}},
  \bibinfo{author}{\bibfnamefont{G.}~\bibnamefont{Trucks}},
  \bibinfo{author}{\bibfnamefont{H.}~\bibnamefont{Schlegel}},
  \bibinfo{author}{\bibfnamefont{G.}~\bibnamefont{Scuseria}},
  \bibinfo{author}{\bibfnamefont{M.}~\bibnamefont{Robb}},
  \bibinfo{author}{\bibfnamefont{J.}~\bibnamefont{Cheeseman}},
  \bibinfo{author}{\bibfnamefont{J.}~\bibnamefont{Montgomery}},
  \bibinfo{author}{\bibfnamefont{T.}~\bibnamefont{Vreven}},
  \bibinfo{author}{\bibfnamefont{K.}~\bibnamefont{Kudin}},
  \bibinfo{author}{\bibfnamefont{J.}~\bibnamefont{Burant}},
  \bibnamefont{et~al.}, \emph{\bibinfo{title}{Gaussian 03, \uppercase{R}evision
  \uppercase{C}.02}} (\bibinfo{year}{2004}),
  \bibinfo{note}{\uppercase{G}aussian, Inc., Wallingford, CT}.

\bibitem[{\citenamefont{Darden et~al.}(1993)\citenamefont{Darden, York, and
  Pedersen}}]{darden1993particle}
\bibinfo{author}{\bibfnamefont{T.}~\bibnamefont{Darden}},
  \bibinfo{author}{\bibfnamefont{D.}~\bibnamefont{York}}, \bibnamefont{and}
  \bibinfo{author}{\bibfnamefont{L.}~\bibnamefont{Pedersen}},
  \bibinfo{journal}{Journal of Chemical Physics} \textbf{\bibinfo{volume}{98}},
  \bibinfo{pages}{10089} (\bibinfo{year}{1993}).

\bibitem[{\citenamefont{Hess et~al.}(1997)\citenamefont{Hess, Bekker,
  Berendsen, and Fraaije}}]{hess1997lincs}
\bibinfo{author}{\bibfnamefont{B.}~\bibnamefont{Hess}},
  \bibinfo{author}{\bibfnamefont{H.}~\bibnamefont{Bekker}},
  \bibinfo{author}{\bibfnamefont{H.~J.~C.} \bibnamefont{Berendsen}},
  \bibnamefont{and} \bibinfo{author}{\bibfnamefont{J.~G. E.~M.}
  \bibnamefont{Fraaije}}, \bibinfo{journal}{Journal of Computational Chemistry}
  \textbf{\bibinfo{volume}{18}}, \bibinfo{pages}{1463} (\bibinfo{year}{1997}).

\bibitem[{\citenamefont{Jorgensen et~al.}(1983)\citenamefont{Jorgensen,
  Chandrasekhar, Madura, Impey, and Klein}}]{jorgensen1983comparison}
\bibinfo{author}{\bibfnamefont{W.~L.} \bibnamefont{Jorgensen}},
  \bibinfo{author}{\bibfnamefont{J.}~\bibnamefont{Chandrasekhar}},
  \bibinfo{author}{\bibfnamefont{J.~D.} \bibnamefont{Madura}},
  \bibinfo{author}{\bibfnamefont{R.~W.} \bibnamefont{Impey}}, \bibnamefont{and}
  \bibinfo{author}{\bibfnamefont{M.~L.} \bibnamefont{Klein}},
  \bibinfo{journal}{Journal of Chemical Physics} \textbf{\bibinfo{volume}{79}},
  \bibinfo{pages}{926} (\bibinfo{year}{1983}).

\bibitem[{\citenamefont{Alln{\'e}r et~al.}(2012)\citenamefont{Alln{\'e}r,
  Nilsson, and Villa}}]{allner2012magnesium}
\bibinfo{author}{\bibfnamefont{O.}~\bibnamefont{Alln{\'e}r}},
  \bibinfo{author}{\bibfnamefont{L.}~\bibnamefont{Nilsson}}, \bibnamefont{and}
  \bibinfo{author}{\bibfnamefont{A.}~\bibnamefont{Villa}},
  \bibinfo{journal}{Journal of Chemical Theory and Computation}
  \textbf{\bibinfo{volume}{8}}, \bibinfo{pages}{1493} (\bibinfo{year}{2012}).

\bibitem[{\citenamefont{Bussi et~al.}(2007)\citenamefont{Bussi, Donadio, and
  Parrinello}}]{bussi2007canonical}
\bibinfo{author}{\bibfnamefont{G.}~\bibnamefont{Bussi}},
  \bibinfo{author}{\bibfnamefont{D.}~\bibnamefont{Donadio}}, \bibnamefont{and}
  \bibinfo{author}{\bibfnamefont{M.}~\bibnamefont{Parrinello}},
  \bibinfo{journal}{Journal of Chemical Physics}
  \textbf{\bibinfo{volume}{126}}, \bibinfo{pages}{014101}
  (\bibinfo{year}{2007}).

\bibitem[{\citenamefont{Berendsen et~al.}(1984)\citenamefont{Berendsen, Postma,
  van Gunsteren, DiNola, and Haak}}]{berendsen1984molecular}
\bibinfo{author}{\bibfnamefont{H.~J.~C.} \bibnamefont{Berendsen}},
  \bibinfo{author}{\bibfnamefont{J.~P.~M.} \bibnamefont{Postma}},
  \bibinfo{author}{\bibfnamefont{W.~F.} \bibnamefont{van Gunsteren}},
  \bibinfo{author}{\bibfnamefont{A.}~\bibnamefont{DiNola}}, \bibnamefont{and}
  \bibinfo{author}{\bibfnamefont{J.~R.} \bibnamefont{Haak}},
  \bibinfo{journal}{Journal of Chemical Physics} \textbf{\bibinfo{volume}{81}},
  \bibinfo{pages}{3684} (\bibinfo{year}{1984}).

\bibitem[{\citenamefont{Grossfield}()}]{gross209}
\bibinfo{author}{\bibfnamefont{A.}~\bibnamefont{Grossfield}},
  \emph{\bibinfo{title}{Wham: an implementation of the weighted histogram
  analysis method. version 2.0.9}},
  \urlprefix\url{http://membrane.urmc.rochester.edu/content/wham/}.

\bibitem[{\citenamefont{Klein et~al.}(2004)\citenamefont{Klein, Moore, and
  Steitz}}]{klein2004roles}
\bibinfo{author}{\bibfnamefont{D.}~\bibnamefont{Klein}},
  \bibinfo{author}{\bibfnamefont{P.}~\bibnamefont{Moore}}, \bibnamefont{and}
  \bibinfo{author}{\bibfnamefont{T.}~\bibnamefont{Steitz}},
  \bibinfo{journal}{Journal of Molecular Biology}
  \textbf{\bibinfo{volume}{340}}, \bibinfo{pages}{141} (\bibinfo{year}{2004}).

\bibitem[{\citenamefont{Gendron et~al.}(2001)\citenamefont{Gendron, Lemieux,
  and Major}}]{gendron2001quantitative}
\bibinfo{author}{\bibfnamefont{P.}~\bibnamefont{Gendron}},
  \bibinfo{author}{\bibfnamefont{S.}~\bibnamefont{Lemieux}}, \bibnamefont{and}
  \bibinfo{author}{\bibfnamefont{F.}~\bibnamefont{Major}},
  \bibinfo{journal}{Journal of Molecular Biology}
  \textbf{\bibinfo{volume}{308}}, \bibinfo{pages}{919} (\bibinfo{year}{2001}).

\bibitem[{\citenamefont{Bottaro et~al.}(2014)\citenamefont{Bottaro, Di~Palma,
  and Bussi}}]{bottaro14nar}
\bibinfo{author}{\bibfnamefont{S.}~\bibnamefont{Bottaro}},
  \bibinfo{author}{\bibfnamefont{F.}~\bibnamefont{Di~Palma}}, \bibnamefont{and}
  \bibinfo{author}{\bibfnamefont{G.}~\bibnamefont{Bussi}},
  \bibinfo{journal}{Nucleic Acids Research}
  \textbf{\bibinfo{volume}{42}}, \bibinfo{pages}{13306} (\bibinfo{year}{2014}).

\bibitem[{\citenamefont{Salim et~al.}(2012)\citenamefont{Salim, Lamichhane,
  Zhao, Banerjee, Philip, Rueda, and Feig}}]{salim2012thermodynamic}
\bibinfo{author}{\bibfnamefont{N.}~\bibnamefont{Salim}},
  \bibinfo{author}{\bibfnamefont{R.}~\bibnamefont{Lamichhane}},
  \bibinfo{author}{\bibfnamefont{R.}~\bibnamefont{Zhao}},
  \bibinfo{author}{\bibfnamefont{T.}~\bibnamefont{Banerjee}},
  \bibinfo{author}{\bibfnamefont{J.}~\bibnamefont{Philip}},
  \bibinfo{author}{\bibfnamefont{D.}~\bibnamefont{Rueda}}, \bibnamefont{and}
  \bibinfo{author}{\bibfnamefont{A.~L.} \bibnamefont{Feig}},
  \bibinfo{journal}{Biophysical journal} \textbf{\bibinfo{volume}{102}},
  \bibinfo{pages}{1097} (\bibinfo{year}{2012}).

\bibitem[{\citenamefont{Stephenson et~al.}(2013)\citenamefont{Stephenson,
  Asare-Okai, Chen, Keller, Santiago, Tenenbaum, Garcia, Fabris, and
  Li}}]{stephenson2013essential}
\bibinfo{author}{\bibfnamefont{W.}~\bibnamefont{Stephenson}},
  \bibinfo{author}{\bibfnamefont{P.~N.} \bibnamefont{Asare-Okai}},
  \bibinfo{author}{\bibfnamefont{A.~A.} \bibnamefont{Chen}},
  \bibinfo{author}{\bibfnamefont{S.}~\bibnamefont{Keller}},
  \bibinfo{author}{\bibfnamefont{R.}~\bibnamefont{Santiago}},
  \bibinfo{author}{\bibfnamefont{S.~A.} \bibnamefont{Tenenbaum}},
  \bibinfo{author}{\bibfnamefont{A.~E.} \bibnamefont{Garcia}},
  \bibinfo{author}{\bibfnamefont{D.}~\bibnamefont{Fabris}}, \bibnamefont{and}
  \bibinfo{author}{\bibfnamefont{P.~T.} \bibnamefont{Li}},
  \bibinfo{journal}{Journal of the American Chemical Society}
  \textbf{\bibinfo{volume}{135}}, \bibinfo{pages}{5602} (\bibinfo{year}{2013}).

\bibitem[{\citenamefont{Ban{\'a}\v{s} et~al.}(2012)\citenamefont{Ban{\'a}\v{s},
  Ml{\'a}dek, Otyepka, Zgarbov{\'a}, Jure\v{c}ka, Svozil, Lanka\v{s}, and
  \v{S}poner}}]{banas2012can}
\bibinfo{author}{\bibfnamefont{P.}~\bibnamefont{Ban{\'a}\v{s}}},
  \bibinfo{author}{\bibfnamefont{A.}~\bibnamefont{Ml{\'a}dek}},
  \bibinfo{author}{\bibfnamefont{M.}~\bibnamefont{Otyepka}},
  \bibinfo{author}{\bibfnamefont{M.}~\bibnamefont{Zgarbov{\'a}}},
  \bibinfo{author}{\bibfnamefont{P.}~\bibnamefont{Jure\v{c}ka}},
  \bibinfo{author}{\bibfnamefont{D.}~\bibnamefont{Svozil}},
  \bibinfo{author}{\bibfnamefont{F.}~\bibnamefont{Lanka\v{s}}},
  \bibnamefont{and}
  \bibinfo{author}{\bibfnamefont{J.}~\bibnamefont{\v{S}poner}},
  \bibinfo{journal}{Journal of Chemical Theory and Computation}
  \textbf{\bibinfo{volume}{8}}, \bibinfo{pages}{2448} (\bibinfo{year}{2012}).

\bibitem[{\citenamefont{Grubmuller et~al.}(1996)\citenamefont{Grubmuller,
  Heymann, and Tavan}}]{grubmuller1996ligand}
\bibinfo{author}{\bibfnamefont{H.}~\bibnamefont{Grubmuller}},
  \bibinfo{author}{\bibfnamefont{B.}~\bibnamefont{Heymann}}, \bibnamefont{and}
  \bibinfo{author}{\bibfnamefont{P.}~\bibnamefont{Tavan}},
  \bibinfo{journal}{Science} \textbf{\bibinfo{volume}{271}},
  \bibinfo{pages}{997} (\bibinfo{year}{1996}).

\bibitem[{\citenamefont{Minh and Adib}(2008)}]{minh2008optimized}
\bibinfo{author}{\bibfnamefont{D.~D.} \bibnamefont{Minh}} \bibnamefont{and}
  \bibinfo{author}{\bibfnamefont{A.~B.} \bibnamefont{Adib}},
  \bibinfo{journal}{Physical Review Letters} \textbf{\bibinfo{volume}{100}},
  \bibinfo{pages}{180602} (\bibinfo{year}{2008}).

\bibitem[{\citenamefont{Do et~al.}(2013)\citenamefont{Do, Carloni, Varani, and
  Bussi}}]{do2013rna}
\bibinfo{author}{\bibfnamefont{T.~N.} \bibnamefont{Do}},
  \bibinfo{author}{\bibfnamefont{P.}~\bibnamefont{Carloni}},
  \bibinfo{author}{\bibfnamefont{G.}~\bibnamefont{Varani}}, \bibnamefont{and}
  \bibinfo{author}{\bibfnamefont{G.}~\bibnamefont{Bussi}},
  \bibinfo{journal}{Journal of Chemical Theory and Computation}
  \textbf{\bibinfo{volume}{9}}, \bibinfo{pages}{1720} (\bibinfo{year}{2013}).

\end{thebibliography}

%
%
%
%

%
%
%
%
%
%
%
%
%
%

%
%

\end{document}